\documentclass[%
 reprint,
 amsmath,amssymb,
 aps,
]{revtex4-2}

\usepackage{graphicx}
\usepackage{dcolumn}
\usepackage{bm}

\usepackage{braket}
\usepackage[colorlinks,linkcolor=blue,anchorcolor=blue,citecolor=blue,urlcolor=blue]{hyperref}

\begin{document}

\preprint{APS/123-QED}

\title{Dynamics of string breaking and revival in a Rydberg atomic chain}

\author{Xin Liu$^{1,2}$}
\thanks{These authors contributed equally to this work.}
\author{Han-Chao Chen$^{1,2}$}
\thanks{These authors contributed equally to this work.}
\author{Zheng-Yuan Zhang$^{1,2}$}
\thanks{These authors contributed equally to this work.}
\author{Jun Zhang$^{1,2}$}
\author{Ya-Jun Wang$^{1,2}$}
\author{Qing Li$^{1,2}$}
\author{Shi-Yao Shao$^{1,2}$}
\author{Bang Liu$^{1,2}$}
\author{Li-Hua Zhang$^{1,2}$}
\author{Dong-Sheng Ding$^{1,2}$}
\email{dds@ustc.edu.cn}
\author{Bao-Sen Shi$^{1,2}$}

\affiliation{$^1$Laboratory of Quantum Information, University of Science and Technology of China, Hefei 230026, China.}
\affiliation{$^2$Anhui Province Key Laboratory of Quantum Network, University of Science and Technology of China, Hefei 230026, China.}

\date{\today}

\begin{abstract}
String breaking is one of the most representative non-perturbative dynamics processes in confinement theory, typically associated with the creation of particle–antiparticle pairs. In this paper, we take a one-dimensional Rydberg atomic chain to theoretically study the dynamical of finite-length string state. Under different string tension conditions, we find that the string dynamics exhibits two clearly distinguishable evolution characteristics: one is that the string breaks and the system enters a superposition state space containing multiple meson state configurations; the other is localized string dynamics, in which the string undergoes local breaking but can then recombine and return to a state close to the initial structure, with the breaking and recombination processes recurring over a long time scale. Through the analysis of the evolution of different meson state configurations, we visually depict the redistribution of configuration weights during the string breaking process, and reveal the observable recovery characteristics of the string after breaking. Further analysis shows that the enhancement of quantum fluctuations can increase the weight of the double-meson state configurations in the system wave function without changing the dominant dynamical behavior. The above results present a rich picture of string breaking dynamics in a one-dimensional Rydberg atomic chain and provide insights for studying confinement physics and related gauge field theory phenomena on quantum simulation platforms. 
\end{abstract}

\maketitle


\section{introduction}
The confinement mechanism is a fundamental feature in quantum chromodynamics: the static quark-antiquark potential energy between color charges approximately linearly increases with distance \cite{1,2,3,4}. This behavior is usually described as the formation of color flux tubes or strings connecting quark pairs \cite{5,6,7,8}. In the presence of dynamical matter fields, when the accumulated energy is sufficient to generate new pair excitations, the original string state will become unstable and may break, corresponding to string breaking \cite{9,10,11,12,13,14,15,16,17,18}. Lattice gauge theory provides a theoretical framework that is closer to the original confinement problem for studying the dynamics of string breaking. In one-dimensional $\mathrm{U(1)}$ lattice gauge theory (such as the Schwinger model), the generation mechanism and time evolution of string breaking have also been explored through numerical methods \cite{29,30,31,32,33,34}. 

With the development of quantum many-body systems and quantum simulation platforms, the study of string breaking has gradually expanded from the high-energy physics context to a series of low-dimensional, controllable model systems \cite{19,20,21,22,23,24}. In these systems, although there are no real color charge degrees of freedom, through the design of effective interactions or normative constraints, physical behaviors highly similar to the confinement phenomenon can be achieved in the low-energy subspace. At the same time, theoretical proposals and experimental progress based on quantum simulation platforms such as cold atoms, ion traps, and Rydberg atomic arrays have opened up new avenues for directly observing and manipulating the dynamics related to confinement \cite{35,36,37,38,39,40,41,42,56,57}. For instance, in one-dimensional spin chains or spin-fermion models, introducing long-range or linearly growing effective potentials can induce string-like structures between domain walls, whose breaking process is manifested as the generation of pairs of excitations and their propagation over time. The related dynamical features have been systematically studied \cite{19,20,25,26,27,28}. In the $(2+1)$-dimensional programmable Rydberg atom quantum simulator, a linear confinement potential is achieved by regulating the atomic arrangement and local detuning, and the string breaking process and its dynamical evolution have been directly observed \cite{41}. Moreover, theoretical studies have shown that the Rydberg atom system can simulate the gauge field dynamics in one-dimensional quantum electrodynamics, including string structures and the generation and propagation of particle–antiparticle pairs \cite{22}. However, most of the existing studies focus on the mechanism of string breaking and its early dynamic behavior. It is interesting to investigate the subsequent evolution of the system after the fracture, such as free evolution or constrained dynamics. 

\begin{figure*}[ht]
    \centering
    \includegraphics[width=1.0\linewidth]{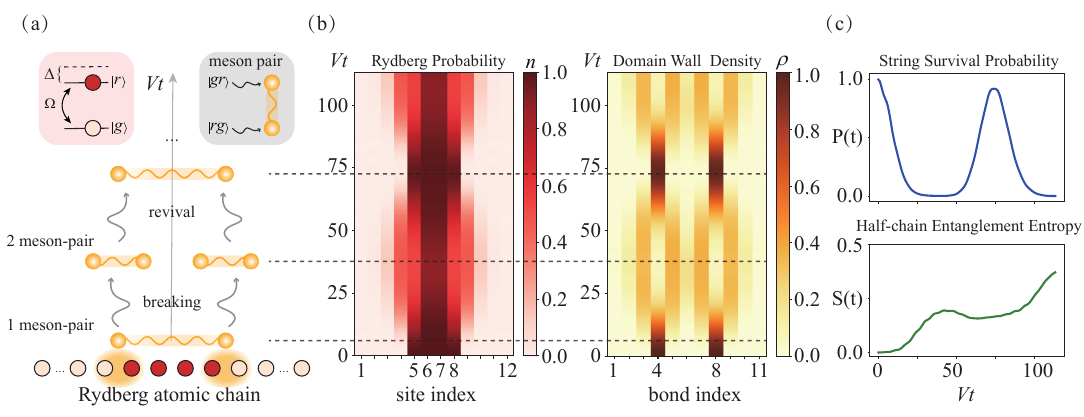}
    \caption{\textbf{The string-meson physical picture and dynamics in a Rydberg atomic chain. }(a) Schematic diagram of a one-dimensional Rydberg atomic chain and the corresponding string-meson mapping. Each atom is modeled as a two-level system consisting of the ground state $\ket{g}$ and the Rydberg excited state $\ket{r}$, coherently coupled by a laser field with Rabi frequency $\Omega$ and detuning $\Delta$. Configurations $\ket{gr}$ and $\ket{rg}$ correspond to domain-wall excitations and are mapped to particle–antiparticle pairs. A contiguous region of Rydberg excitations bounded by two domain walls is mapped to a string. During the dynamical evolution, the initial single-meson state undergoes string breaking, generating two mesons, which subsequently recombine, in correspondence with the numerical results shown in panels (b) and (c). (b) Time evolution of the Rydberg excitation probability $n_{i}(t)$ and the domain wall density $D_{i}(t)$ as heat maps versus the dimensionless time $Vt$. The horizontal axes correspond to the atomic site index and the bond index, respectively, while the vertical axis represents the dimensionless time $Vt$, and the color scale indicates the magnitude of the domain wall density. (c) Time evolution of the string survival probability $P(t)$ and the half-chain entanglement entropy $S(t)$ as functions of $Vt$. }
    \label{fig:System}
\end{figure*}

In this paper, we use a one-dimensional Rydberg atomic chain to construct a mapping model between atomic direct-product states and meson states. We propose a dynamical interpretation framework based on the meson subspace to analyze the breaking and recovery behaviors of finite-length string states during the evolution process from the perspective of configuration space. We interpret string breaking as the dynamical diffusion of the probability amplitude of the quantum state from the initial string configuration to different meson configurations under the string-meson configuration basis, and the recovery behavior corresponds to the rebound of the probability of the initial single-meson state during the evolution process. In addition, we studied the influence of string tension and quantum fluctuation changes on the distribution characteristics of different meson configurations. 

\section{Theoretical model and mapping}
We study a one-dimensional Rydberg atomic chain with open boundary conditions. The system consists of $L$ equally spaced atoms, where each atom $i$ is a two-level system with a ground state $\ket{g_{i}} \equiv \ket{0_{i}}$ and a Rydberg excited state $\ket{r_{i}} \equiv \ket{1_{i}}$. All atoms are driven by a uniform optical field, and under the rotation wave approximation, the effective Hamiltonian of the system can be written as \cite{43,44,45,46,47,48,49,50}:
\begin{align}
\hat{H} = \frac{\Omega}{2}\sum_{i}^{L}(\ket{r_i}\bra{g_i} + \ket{g_i}\bra{r_i}) - \Delta\sum_{i}^{L}n_i + V\sum_{i}^{L-1}n_i n_{i+1}
\end{align}
where $\Omega$ is the laser-induced Rabi frequency, and $\Delta$ is the laser detuning from the atomic transition. The operator $\hat{n}_i = \ket{r_i}\bra{r_i}$ is the occupation number operator that projects the 
$i$-th atom onto the Rydberg excited state. The interaction term $V\sum_{i}^{L-1}n_i n_{i+1}$ describes the interaction energy when two neighboring atoms are simultaneously excited to the Rydberg state. This interaction originates from the van der Waals interaction between Rydberg atoms, which takes the form $ V_{ij}=C_{6}/r^{6}_{ij}$ \cite{51,52}, where $C_{6}$ is a constant determined by the atomic species and the chosen Rydberg level, and $r_{ij}$ is the distance between the $i$-th and $j$-th atoms. In this work, the nearest-neighbor interaction strength $V$ is kept fixed and set to $V=2\pi\times12~\mathrm{MHz}$, a typical experimental value for $^{87}\text{Rb}$ atoms \cite{53,54}. This choice sets the energy scale of the system, with the corresponding unit of time given by $V^{-1}$. Accordingly, all dynamical processes discussed in this work are characterized in terms of the dimensionless time $Vt$. Owing to the rapid decay of the van der Waals interaction with distance, next-nearest-neighbor and longer-range interactions can be safely neglected; therefore, only nearest-neighbor interactions are retained throughout this work \cite{48,51}.

In a Rydberg atomic chain, the spatial distribution of Rydberg excitations is jointly determined by the laser detuning and the strength of interatomic interactions. When the detuning satisfies the condition $\Delta=V$, the system enters the anti-blockade regime. With the large detuning, the coupling between the ground state and Rydberg state of an isolated atom is strongly suppressed, making its excitation energetically unfavorable. However, if an atom is already in the Rydberg excited state, the effective detuning for exciting its nearest neighbor from $\ket{g}$ to $\ket{r}$ becomes $\Delta-V$. When $\Delta=V$, this excitation process is energetically resonant, and the probability of generating a Rydberg excitation is therefore greatly enhanced. As a result, this anti-blockade mechanism causes local excitations to promote the excitation of nearby atoms into the Rydberg state, leading to the formation of contiguous excited regions during the dynamical evolution. Each contiguous region of identical states can be regarded as a domain. When the $i$-th and $(i+1)$-th atoms belong to different domains, a domain wall is defined between them. The corresponding domain-wall density operator is defined as $D_{i}(t)=(\hat{I}-\hat{\sigma}^z_{i}(t)\hat{\sigma}^z_{i+1}(t))/2$, where $\hat{\sigma}^z_{i}=\ket{r_i}\bra{r_i}-\ket{g_i}\bra{g_i}$. 

Under open boundary conditions, the creation and annihilation of domain walls necessarily occur in pairs, and a single domain wall cannot stably exist as an independent low-energy excitation \cite{55}. We consider a region consisting of consecutive Rydberg excitations $\ket{r}$, this is bounded by two domain walls on its left and right. We define the length of this region as $l$, corresponding to the number of atoms in the Rydberg excited state within the region. For a given $l$, the energy of this region mainly arises from the difference between the detuning term and the interaction between neighboring Rydberg excitations, and can be written as $\Delta E(l)=(l-1)(V-\Delta)$. When $V\ne\Delta$, an effective interaction potential between the two domain walls emerges, which grows linearly with their separation. This provides a confinement mechanism between domain walls, causing paired domain walls to naturally form bound structures during the dynamical evolution, analogous to particle–antiparticle pairs confined in gauge field theories. Accordingly, we map a pair of domain walls together with the region between them onto a confined meson, as illustrated in Fig.~\ref{fig:System}(a). The contiguous region of Rydberg excitations between the two domain walls can be naturally mapped onto a string, whose length is determined by the separation between the domain walls. The energy cost of the string originates from the difference between the detuning term and the nearest-neighbor Rydberg interaction, thereby generating an effective confining potential between the domain walls, often described in terms of an effective string tension. Within this physical picture, string breaking and recombination correspond to the creation and annihilation of domain wall pairs, which in turn represents the creation and annihilation of mesons.  

\section{String breaking and revival dynamics in a one-dimensional Rydberg atomic chain}
Within the above string–meson picture, we investigate the dynamics of finite-length string states. In this work, we consider a one-dimensional Rydberg atomic chain with open boundary conditions and a total of $L=12$ atoms. The initial state of the system is defined as
\begin{align}
\ket{\psi_{\text{string}}}=\ket{g}^{\otimes4}\otimes\ket{r}^{\otimes4}\otimes\ket{g}^{\otimes4},
\end{align}
which represents a contiguous Rydberg-excited region of length $l=4$ located at the center of the chain, with all other atoms remaining in the ground state. According to the excitation–meson mapping discussed in the Model section, this composite object consisting of two domain walls and the contiguous Rydberg excited region between them can be regarded as a single confined meson. Equivalently, this product state corresponds to a finite-length string state.

\begin{figure*}[ht]
    \centering
    \includegraphics[width=1.0\linewidth]{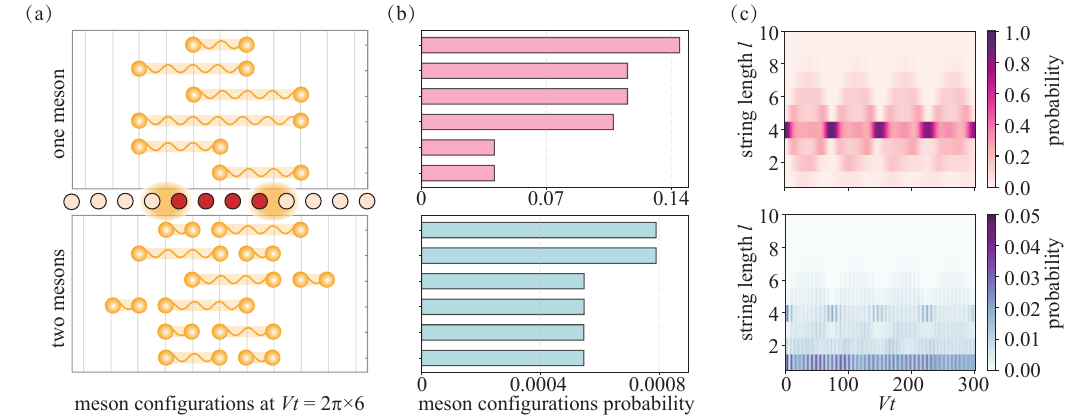}
    \caption{\textbf{Meson configurations and string length dynamics during the string breaking process. }(a) Examples of the main single-meson (top) and double-meson (bottom) configurations that the system can evolve into from the initial single-meson state with string length $l = 4$ at $Vt = 2\pi\times6$. (b) The occurrence probabilities of various configurations in (a). (c) Time evolution of string length distribution in the single-meson (top) and double-meson (bottom) subspaces with $Vt$. The horizontal axis represents time $Vt$, the vertical axis represents string length $l$, and the color bar indicates the probability of occurrence of that string length.}
    \label{fig:Meson}
\end{figure*}

We first choose the detuning parameter $\Delta=2\pi\times11~\mathrm{MHz}$ to analyze the dynamics of the finite-length string in the off-resonant regime. In this case, $\Delta E(l)>0$, indicating the presence of a repulsive confining potential within the meson. As shown in Fig.~\ref{fig:System}(b), the left panel depicts the time evolution of the Rydberg occupation probability at each lattice site. As time evolves, the Rydberg occupation probability spreads from this contiguous region to neighboring sites. Around $Vt\approx50$, the distribution flows back toward the center of the chain, and the system returns to a configuration close to the initial Rydberg excitation pattern. The right panel of Fig.~\ref{fig:System}(b) shows the corresponding dynamics of the domain wall density. Initially, domain walls exist only on bonds $4$ and $8$, corresponding to the two boundaries of the contiguous Rydberg excitation region. As time evolves, the domain wall density on these bonds spreads to neighboring bonds, with pronounced distributions appearing on bonds $3$, $5$, and $7$, $9$, while only weak distributions are present on other bonds. At later times, the spatial distribution of the domain wall density gradually converges, becoming localized again. This revival behavior indicates that the domain walls do not irreversibly spread across the entire chain, but instead remain confined within a relatively localized region and undergo recurrent dynamics over time.

Within the framework of the string mapping, the dynamical evolution of the meson state can be described by the simplified image in Fig.~\ref{fig:System}(a). In the evolution process, a single-meson state can transform into a double-meson state, that is, the single segment string structure corresponding to a continuous excitation region splits, forming two separate strings. The process from the single-meson state to the double-meson state implies the existence of a string breaking process. Since the energy of the continuous $\ket{r}$ excitation region accumulates as its length increases, a longer excitation region means that the system carries higher energy during the evolution. There is an energy difference between different meson state configurations, and under the action of quantum fluctuations, a transformation from the single-meson state to the double-meson state is allowed. In a Rydberg atomic chain, that is, a pair of new domain walls are generated within the original excitation region. The newly generated domain walls act as breaking points, dividing the single continuous $\ket{r}$ region into two segments and separating them by the $\ket{g}$ background. The double-meson state configuration does not remain unchanged once generated. Over time, under the continuous action of quantum fluctuations, there is a probability that the two separated excitation regions will recombine into one, corresponding to the dynamical recovery of the double-meson state to the single-meson state. The entire process of the evolution path is that a finite-length string undergoes temporary breaking and recovery during the dynamical process.

To quantitatively characterize the dynamical behavior of this finite-length string during time evolution, we first define the string survival probability
\begin{align}
P(t)=\vert\langle{\psi_{\text{string}}}\ket{\psi(t)}\vert^{2},
\end{align}
which measures the probability for the system to return to the initial string state at time $t$. In addition, we divide the system into two equal subsystems, $A$ and $B$, and define the half-chain entanglement entropy as
\begin{align}
S(t)=-\mathrm{Tr}[\rho_A(t)\ln\rho_A(t)],
\end{align}
where $\rho_A(t)=\mathrm{Tr_B}\ket{\psi(t)}\bra{\psi(t)}$ is the reduced density matrix obtained by tracing over subsystem $B$. This quantity can be used to characterize the entanglement generation across the interface between subsystems $A$ and $B$ during the dynamical process. 

Figure.~\ref{fig:System}(c) shows the evolution of the string survival probability $P(t)$ and the half-chain entanglement entropy $S(t)$ over time. During the early stage of the evolution, $P(t)$ decreased significantly from its initial value of $1$, indicating an effective coupling between the initial atomic chain direct product state and other accessible configurations, including the process of generating additional domain wall pairs, corresponding to string breaking. Subsequently, $P(t)$ does not monotonically decay but instead exhibits distinct quasi-periodic oscillations. At certain moments, $P(t)$ recovers to a value close to $1$, indicating that the system state periodically gains a large overlap with the initial single-meson state during its evolution, rather than irreversibly moving away from the initial chord state. This quasi-periodic revival can be understood as follows: Under coherent driving, meson configurations with the same energy can freely interconvert, leading to coherent transitions among degenerate states. For meson states with finite energy differences, the coherent drive provides quantum fluctuations that allow the system to overcome these energy mismatches, enabling probabilistic transitions between such states. As a result, the system can undergo transitions among different meson configurations after string breaking and eventually return close to the initial state. Starting from a single-meson state, the system can evolve along multiple dynamical channels to other meson configurations, and then return to the initial state at different characteristic time scales later. Due to the inconsistent evolution periods of each channel, the system state has a relatively small overlap with the initial state most of the time. However, at certain time scales that satisfy the common multiple, the system state as a whole approaches the initial single-meson state again, giving rise to quasi-periodic revivals observed in the curve of the string survival probability $P(t)$. 

\begin{figure*}[ht]
    \centering
    \includegraphics[width=1.0\linewidth]{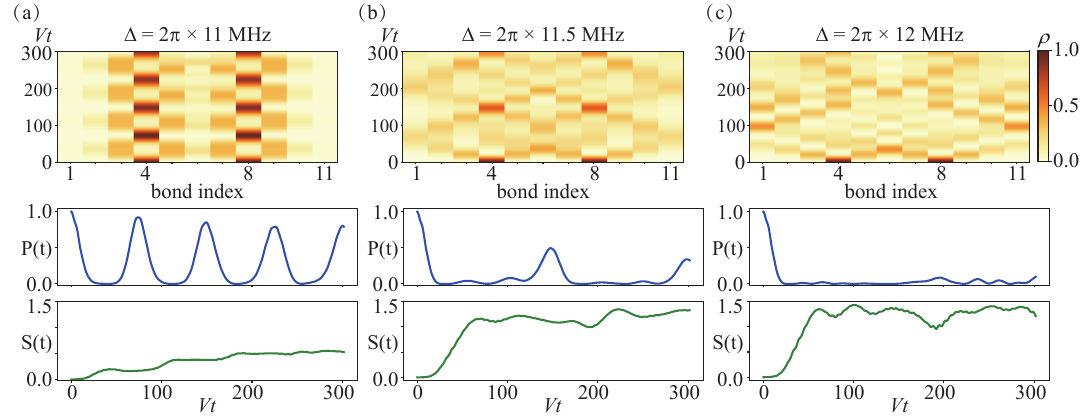}
    \caption{\textbf{The dynamical evolution of finite-length string states (initial string length $l=4$) under different detuning parameters.} In Figs (a)-(c), the dynamical results of the system are presented for fixed interaction strength $V$ and detuning parameters $\Delta=2\pi\times11~\mathrm{MHz}$ (a), $\Delta=2\pi\times11.5~\mathrm{MHz}$ (b),and $\Delta=2\pi\times12~\mathrm{MHz}$ (c), respectively. In each column, from top to bottom, the evolution heat map of domain wall density on each bond of the atomic chain with time $Vt$, the string survival probability $P(t)$, and the half-chain entanglement entropy $S(t)$ with time $Vt$ are displayed. Under different detuning parameters, the three dynamical observables exhibit significantly different time evolution characteristics.}
    \label{fig:Detuning}
\end{figure*}

Correspondingly, the time evolution of the half-chain entanglement entropy $S(t)$ provides complementary information about the dynamics. While $P(t)$ only quantifies the projection weight of the system onto the initial state, the wave function $\psi(t)$ may still contain finite amplitudes distributed over other configurations, even during the time windows when $P(t)$ exhibits a revival. If these non-initial components span the half-chain division surface, they will introduce additional entanglement contributions. $S(t)$ shows an overall trend of slow growth over time, indicating that entanglement gradually accumulates during the evolution. Therefore, the overlap of the system state $\psi(t)$ with the initial state is difficult to repeatedly return to $P(t) = 1$ over a long period of time, meaning there is no unlimited revival. On the other hand, $S(t)$ does not increase monotonically, but shows a relative decrease during the time intervals when $P(t)$ revives, indicating a temporary reduction of entanglement across the bipartition. In these intervals, the system state is more strongly dominated by the initial single-meson component, consistent with the revival behavior observed in $P(t)$. 

By analyzing the evolution of single-meson states and double-meson states and their weight changes, we can directly depict the manifestation of the string breaking and recovery process in the meson representation. We performed projection analysis on the meson states in the single-meson subspace and the double-meson subspace, and the relevant results are shown in Fig.~\ref{fig:Meson}. Different meson states can be expressed in the form of $\ket{g}^{\otimes{n}}\otimes\ket{r}^{\otimes{m}}\otimes\ket{g}^{\otimes{n}}$, where $m,n\ge1$, and $m$ represents the string length (i.e., the length of the continuous $\ket{r}$ excitation region). Figure.~\ref{fig:Meson}(a) shows the top six single-meson and double-meson configurations with the highest probability weights during the system evolution when $Vt = 2\pi\times6$. Correspondingly, Fig.~\ref{fig:Meson}(b) presents the occurrence probabilities of these configurations in the system wave function, clearly indicating that the single-meson state dominates the weight throughout the evolution process, while the double-meson state only appears with a relatively small probability. The differences between different meson state configurations mainly manifest in the number of mesons, spatial positions, and string lengths. Meson configurations symmetrically about the chain center have the same probability, such as single-meson configurations $\ket{g}^{\otimes{3}}\otimes\ket{r}^{\otimes{4}}\otimes\ket{g}^{\otimes{5}}$ and $\ket{g}^{\otimes{5}}\otimes\ket{r}^{\otimes{4}}\otimes\ket{g}^{\otimes{3}}$, and double-meson configurations $\ket{g}^{\otimes{4}}\otimes\ket{rg}\otimes\ket{r}^{\otimes{3}}\otimes\ket{g}^{\otimes{3}}$ and $\ket{g}^{\otimes{3}}\otimes\ket{r}^{\otimes{3}}\otimes\ket{gr}\otimes\ket{g}^{\otimes{4}}$. Figure.~\ref{fig:Meson}(c) presents the evolution of the string length distribution in the two subspaces over time. In the single-meson subspace, the string length distribution shows a clear periodic behavior and repeatedly returns to the string length $l = 4$ during the evolution. In the double-meson subspace, configurations with string length $l = 1$ occupy an important proportion, indicating that the probability of generating a short string structure after string breaking is greater. 

\section{Effect of string tension on string dynamics}
After observing the dynamic behavior of the string recovery, we further selected several representative parameter values within the parameter range of $\Delta\leq V$ for numerical calculations. As shown in Fig.~\ref{fig:Detuning}, under the condition of a fixed interaction strength $V$, we compared and studied the dynamic characteristics of the system when the detuning $\Delta$ is $2\pi\times11~\mathrm{MHz}$, $2\pi\times11.5~\mathrm{MHz}$, and $2\pi\times12~\mathrm{MHz}$, and analyzed the differences in the system's dynamics, such as the domain wall dynamics, the survival probability of the string, and the evolution of entanglement. When $\Delta = 2\pi\times11~\mathrm{MHz}$, the domain wall density mainly evolves within a limited range of the chain, and its distribution undergoes diffusion and reflux over time. The main weight of the system wave function is distributed in the single-meson subspace, and it periodically shifts to the double-meson subspace during the time evolution, as shown in Fig.~\ref{fig:Detuning}(a). Correspondingly, the string survival probability $P(t)$ displays clear oscillatory behavior, and the growth of the half-chain entanglement entropy $S(t)$ remains relatively slow. When $\Delta$ increases to $2\pi\times11.5~\mathrm{MHz}$, the domain wall dynamics undergo significant changes, as shown in Fig.~\ref{fig:Detuning}(b). At this time, the energy differences between contiguous $\ket{r}$ excited configurations of different lengths and spatial positions decrease, and the system wave function is more likely to expand between these configurations. According to the relationship between string and mesons mapping, this behavior can be understood as the reduction in the string tension $\sigma=\lvert{V-\Delta}\rvert$, the energy cost of stretching the string decreases, and the system gradually transitions from a strong confinement dynamics to a weak confinement region. Under this parameter, the diffusion degree of the domain wall density increases, and its distribution covers a larger chain interval. Correspondingly, the oscillation period of $P(t)$ lengthens and the amplitude decreases, and the system can still observe the process of the wave function approaching the initial configuration during the evolution, but its characteristic time is significantly prolonged compared to $\Delta = 2\pi\times11~\mathrm{MHz}$, and the recovery fidelity decreases. The growth of half-chain entanglement entropy $S(t)$ is more significant and tends to saturate at higher values, reflecting the continuous accumulation of entanglement across the half-chain separation surface during evolution, and the system enters a high entanglement dynamics state dominated by multiple configuration mixtures.

\begin{figure*}[ht]
    \centering
    \includegraphics[width=1.0\linewidth]{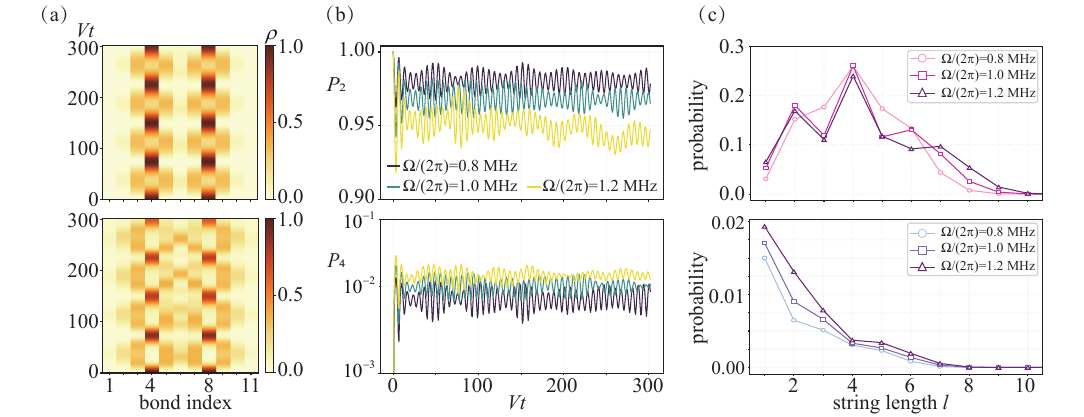}
    \caption{\textbf{The influence of driving intensity on string evolution and meson configurations. }(a) Evolution heat map of domain wall density over time under fixed interaction strength $V$ and detuning parameter $\Delta$. The upper and lower plots correspond to $\Omega/(2\pi) = 0.8~\mathrm{MHz}$ and $\Omega/(2\pi) = 1.2~\mathrm{MHz}$, respectively, for comparing the influence of domain wall fluctuations and spatial distribution under different driving intensities. (b) At a fixed time slice $Vt = 2\pi\times6$, the total probability weights of the system wave function projected onto different meson subspaces. Shown are the total probability $P_2$ of single-meson configurations and the total probability $P_4$ of double-meson configurations as functions of $\Omega$. (c) At the same time slice $Vt = 2\pi\times6$, the string length distribution is statistically analyzed in the single-meson subspaces and double-meson subspaces, showing the influence of the driving intensity on the internal length structure of the configuration.}
    \label{fig:Driving}
\end{figure*}

Specifically, when $\Delta=V=2\pi\times12~\mathrm{MHz}$, the system enters the resonance region, at which the string tension $\sigma=\lvert{V-\Delta}\rvert=0$. Under this condition, the atomic chain configurations with different numbers of domain walls become highly similar in energy, and the generation and propagation of domain wall pairs become active. As shown in Fig.~\ref{fig:Detuning}(c), the density distribution of domain walls on the initial two bonds gradually spreads to multiple bonds of the entire chain during the evolution process and remains in a widely distributed form over a long time scale. Unlike the case where $\Delta < V$, in the resonance condition, the spatial distribution of domain wall density does not re-focus back to the initial two bond positions in the subsequent evolution but continues to exhibit a spatially discrete characteristic, indicating that the recovery of the initial single-meson configuration is suppressed. The string survival probability $P(t)$ rapidly decays from the initial value of $1$ to a value close to $0$ and only shows small amplitude oscillations in the long-term evolution, and the system hardly returns to the initial state. The half-chain entanglement entropy $S(t)$ grows rapidly in the early stage ($Vt < 100$) and then saturates at a higher value (approximately $1.5$) later. Compared with the parameter range where there is string recovery behavior, the entanglement entropy of the system at $\Delta = V$ is higher. 

\section{Effect of quantum fluctuation strength on string dynamics}
Under the conditions of fixed interaction intensity $V=2\pi\times12~\mathrm{MHz}$ and detuning parameter $\Delta=2\pi\times11~\mathrm{MHz}$, the driving intensity $\Omega$ affects the fluctuation characteristics of the number and spatial distribution of domain walls during the evolution. We investigated the influence of different driving intensities $\Omega$ on the dynamics of domain walls. As shown in Fig.~\ref{fig:Driving}(a), as $\Omega$ increases, the probability of domain walls appearing on other bonds during evolution increases. We calculated the weights of the system's wave function in different domain wall number subspaces at the time slice of $Vt = 2\pi\times6$, as shown in Fig.~\ref{fig:Driving}(b). The results indicate that the system's evolution is mainly distributed in the subspace with $N_{\mathrm{DW}} = 2$ domain walls, and the total probability $P_2$ slightly decreases as $\Omega$ increases; while the total probability $P_4$ of the $N_{\mathrm{DW}} = 4$ configurations increases with the increase of $\Omega$. Under the string-meson mapping relationship, the above changes in domain wall dynamics correspond to the redistribution of the weights of single-meson and double-meson configurations. With the enhancement of quantum fluctuations, the total probability of the single-meson state decreases, while the appearance probability of the double-meson state increases, reflecting that stronger quantum fluctuations make the system more likely to evolve to the configuration containing two mesons. Furthermore, we analyzed the changes in the distribution of string lengths in the two subspaces when $Vt = 2\pi\times6$, as shown in Fig.~\ref{fig:Driving}(c). In the single-meson subspace, the distribution of string lengths does not show a significant systematic dependence on $\Omega$, and the internal structure of the single-meson configuration remains stable within the parameters considered; while in the double-meson subspace, the probability of the appearance of short string configurations increases slightly with increasing $\Omega$. Within the considered parameters, the single-meson state always dominates, and the string breaking process only appears as a secondary dynamic channel driven by enhancement.

\begin{figure}[ht]
    \centering
    \includegraphics[width=0.9\linewidth]{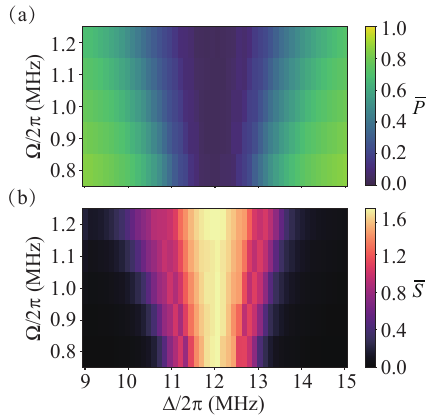}
    \caption{\textbf{The long-term average distribution of string survival probability and half-chain entanglement entropy within the $(\Delta, \Omega)$ parameter space. }(a) The two-dimensional heatmap of the long-term average value $\bar{P}$ of string survival probability in the $(\Delta, \Omega)$ parameter space. (b) The two-dimensional heatmap of the long-term average value $\bar{S}$ of half-chain entanglement entropy in the same parameter space. With $\Omega/(2\pi)\in[0.8, 1.2]~\mathrm{MHz}$, and uniformly selecting $50$ parameter points within the interval of $\Delta/(2\pi)\in[9, 15]~\mathrm{MHz}$ for two-dimensional scanning. The color bar respectively represents the numerical magnitudes of $\bar{P}$ and $\bar{S}$.} 
    \label{fig:Phase}
\end{figure}

To systematically investigate the combined influence of the driving intensity $\Omega$ and the detuning parameter $\Delta$ on the dynamical behavior, we scanned the long-term average values of the string survival probability and the half-chain entanglement entropy in the $(\Delta, \Omega)$ parameter space, as illustrated in Fig.~\ref{fig:Phase}. For any physical quantity $\mathcal{O}(t)$ that evolves over time, its time average is defined as $\mathcal{\bar{O}}(t)=\frac{1}{t_2-t_1}\int_{t_1}^{t_2} \mathcal{O}(t) dt$, where the averaging window is selected after the transient decay in the early stage of system evolution. In this paper, we discard the first $10$ driving cycles and only average the dynamical behavior within the $10th$ to $40th$ cycles, calculating $\bar{P}$ and $\bar{S}$. Figures.~\ref{fig:Phase}(a) and (b) respectively show the distributions of $\bar{P}$ and $\bar{S}$ at different values of $\Delta/(2\pi)$ and $\Omega/(2\pi)$. Both exhibit clear $V$-shaped structures in the parameter space, with lower $\bar{P}$ values in the central part of the parameter space and higher $\bar{P}$ values in the peripheral regions; in contrast, $\bar{S}$ reaches its maximum in the middle of the $V$-shaped structure and significantly decreases in the peripheral regions, demonstrating the complementary relationship between $\bar{P}$ and $\bar{S}$ in the overall trend. 

For configurations such as $\ket{g}^{\otimes{4}}\otimes\ket{r}^{\otimes{4}}\otimes\ket{g}^{\otimes{4}}$ that contain a single continuous Rydberg excitation region, the stretching and translation of the string are mainly driven by the local flipping process at the edge of the excitation region. When $\Delta$ approaches $V$, the edge flipping process approaches an energy resonance, which equivalently corresponds to a reduction in string tension, resulting in a weakening of the energy suppression required for the dynamic processes of changing the string length or shifting the string position. At this time, the system can evolve more freely between different lengths and positions of the single-meson configurations, and further open up dynamic channels leading to a larger configuration set. In this configuration mixing enhancement situation, the projection weight of the wave function on the initial state is significantly diluted, thereby forming a low peak of $\bar{P}$ near $\Delta/(2\pi)\approx12~\mathrm{MHz}$. At the same time, entanglement across the half-chain cut is more easily generated, manifested as an increase in $\bar{S}$. Conversely, when $\lvert{V-\Delta}\rvert$ is relatively large compared to the driving strength $\Omega$, the edge flipping process is inhibited in the dynamics, the probability of changes in string length and position decreases, and the system maintains a larger overlap probability with the initial state during long-time evolution. Therefore, $\bar{P}$ remains high, while the growth of $\bar{S}$ is inhibited. On the other hand, as $\Omega$ increases, quantum fluctuations can overcome larger tension constraints, and the dynamic regions with low $\bar{P}$ and high $\bar{S}$ expand to a larger $\lvert{V-\Delta}\rvert$ range, ultimately forming a $V$-shaped structure in the parameter space that expands with Omega. 

\section{Conclusion and outlook}
This paper reports the dynamics of finite-length string states based on the string-meson mapping relationship in a one-dimensional Rydberg atomic chain. The results of time evolution show that the initial single-meson state undergoes processes of string stretching, breaking, and recombination. Its dynamic characteristics can be characterized by the spatial distribution of domain wall density, the survival probability of the string, and the evolution of the half-chain entanglement entropy. Under different detuning parameters, these observables exhibit significant differences in spatial probability distribution or numerical changes, reflecting the effect of string tension on the system dynamics. Through the projection analysis of the system wave function in different domain wall subspaces, we revealed the evolution characteristics of meson state configurations and string length distribution during string breaking. Over a longer time scale, the system dynamics is always dominated by the single-meson state configuration, while the double-meson state only occupies a relatively small probability weight, but will introduce multi-string configurations, thereby changing the structural characteristics of the string length distribution. With the enhancement of quantum fluctuation intensity, the system can increase the probability of the double-meson state configuration while maintaining the dominance of the single-meson state configuration. Through a two-dimensional scan of the $(\Delta, \Omega)$ parameter space, we have discovered a clear $V$-shaped structure in the long-time average of the string survival probability and the half-chain entanglement entropy: in the central region of the parameter space, the string survival probability decreases while the entanglement entropy reaches a higher level; while away from this region, the system exhibits a more stable string structure and weaker entanglement generation. The research results of this paper provide a feasible theoretical reference for directly observing the dynamical evolution and entanglement characteristics of the string state in experiments.

\begin{acknowledgments}
We acknowledge funding from the National Key R and D Program of China (Grant No. 2022YFA1404002), the National Natural Science Foundation of China (Grant Nos. T2495253, 61525504, 61435011.
\end{acknowledgments}

\nocite{*}

\bibliography{citation}

\end{document}